\documentclass[11pt]{iopart}
\usepackage{iopams}
\usepackage{graphicx}
\usepackage[colorlinks,linkcolor=blue,urlcolor=blue,citecolor=blue]{hyperref}
\usepackage{graphicx}
\usepackage{float}
\usepackage{enumerate}
\newcommand{\D}{{\rm d}}

\def\be{\begin{equation}}
\def\ee{\end{equation}}
\def\bea{\begin{eqnarray}}
\def\eea{\end{eqnarray}}

\begin{document}
\title{Run and tumble  particle  under resetting: a renewal approach}

\date{\today}


\author{Martin R. Evans$^{(1)}$ and Satya N. Majumdar$^{(2)}$}

\address{$^{(1)}$ SUPA, School of Physics and Astronomy, University of
Edinburgh, 
Peter Guthrie Tait Road, Edinburgh EH9 3FD, UK\\
$^{(2)}$ Univ. Paris-Sud, CNRS, LPTMS, UMR 8626, Orsay F-01405, France}

\ead{m.evans@ed.ac.uk, satya.majumdar@u-psud.fr}

\begin{abstract}
We consider a particle undergoing run and tumble dynamics, in which
its velocity stochastically reverses, in one dimension.  We study the
addition of a Poissonian resetting process occurring with rate $r$. At
a reset event the particle's position is returned to the resetting
site $X_r$ and the particle's velocity is reversed with probability
$\eta$.  The case $\eta = 1/2$ corresponds to position resetting and
velocity randomization whereas $\eta =0$ corresponds to position-only
resetting.  We show that, beginning from symmetric initial conditions,
the stationary state does not depend on $\eta$ i.e. it is independent
of the velocity resetting protocol. However, in the presence of an
absorbing boundary at the origin, the survival probability and mean
time to absorption do depend on the velocity 
resetting protocol. Using a renewal equation approach, we show that the 
mean time to absorption is always less for 
velocity randomization than for position-only resetting.
\end{abstract}


\vspace{2pc}

\submitto{To appear in Journal of Physics A: Mathematical and Theoretical 2018}


\section{Introduction}

Stochastic processes  with correlated noise have  a long history in physics  beginning with the Ornstein Uhlenbeck process which generates
a finite correlation time for Brownian motion \cite{UO30,HDL89}. More recently, active matter has been described by stochastic
processes with correlated noise such as the run and tumble dynamics
used to model bacterial motion. Run and tumble dynamics is, in turn,
the continuum limit of persistent random walkers \cite{Weiss,HV10} commonly
used to model animal movement \cite{HLSN00} and search processes
\cite{TVB12}. Also the bidirectional motion of cellular cargoes is, in
general, a correlated random walk \cite{BG13}.

A particle under run and tumble dynamics, in one dimension, obeys a Langevin equation 
of the form
\begin{equation}
\frac{\D x}{\D t} = v_0 \sigma(t)
\label{rnt}
\end{equation}
Here $\sigma$ is  a stochastic process which switches between two states $\sigma = \pm 1$ with rate $\gamma$; non-Poissonian switching has been considered in \cite{Detcheverry}. Thus a run and tumble process is sometimes referred to as `telegraphic noise' to describe the evolution of $\sigma$
\cite{Rosenau}. The  equation for the time evolution of the probability
distribution corresponding to (\ref{rnt})
turns out to be the Telegrapher's equation (see e.g. \cite{Weiss}).
This equation interpolates between the wave equation and the diffusion equation, i.e., when $\gamma \to 0$ we have ballistic motion
and when $\gamma, v_0 \to \infty$ (see section 3.1 equation (\ref{difflim}) for the precise limit) we have diffusive motion.

Run and tumble dynamics  have revealed a number of interesting
nonequilibrium properties such as clustering at boundaries \cite{EG15,Angelani17},
novel stationary states \cite{SEB16, SEB17} and  first-passage properties
\cite{Angelani15,MJKKSMRD18}.
In this paper we
investigate the effect of resetting \cite{EM11a} on run and tumble
dynamics. Resetting is the procedure of restarting a stochastic process from a
given initial condition \cite{MZ99}. It has been shown that resetting profoundly changes the properties of diffusion, a fundamental dynamical process \cite{EM11a}.

Our aim in this work is twofold. First run and tumble particle dynamics allows us to  investigate how the effect of resetting on a process which  interpolates between diffusive and ballistic motion, thus extending previous results which have focussed on diffusion. Second,
a stochastic process of the form (\ref{rnt}) open up a whole new set of
possibilities for resetting, as the velocity variable $\sigma$ as well as the
position $x$ may be reset. Thus the resetting occurs in the phase space of the particle rather than just the position space as has usually been considered. (We note that  orientation resets have  been
considered in continuous time random walks with drifts \cite{MMV17}.)
We focus on the case where the position is reset to a fixed resetting point $X_r$ and 
simultaneously  the  velocity undergoes a resetting protocol. Initially, we consider velocity randomization in which the velocity is  reversed with probability $1/2$. Later we generalise this to the protocol where the velocity is reversed
with probability $\eta$ which allows interpolation to the case of position-only resetting where $\eta=0$.

Let us now summarise recent results on resetting of stochastic processes.
In the case of position resetting,
the resetting position can be fixed to be the  initial condition \cite{EM11a} or chosen from some resetting distribution
\cite{EM11b}. In this way the system is held away from any long time
stationary state and a nonequilibrium stationary state is generated
\cite{EM14}. Interesting transient properties of the relaxation to
this state have been revealed \cite{MSS15}. Also the resetting process
can be considered as a realisation of an intermittent search process
\cite{BenichouRV, MZ02} where a reset event is a long range move.  The resetting
process is usually considered to be a Poisson process with
exponentially distributed waiting times between resetting events,
however more general waiting time distributions including power law
distributions have been considered \cite{EM16,PKE16,NG16}. Moreover,
it has been shown that a deterministic resetting period may be optimal
in the minimisation of mean search times or mean first passage times \cite{PKE16,BDR16,HK16,PR17}.
Resetting with memory, where a walker resets only to previously visited sites
with a certain distribution, have also been 
studied~\cite{BS2014,MSS15b,BEM17,FBGM17}.
While some interesting properties of the mean first-passage time and
its fluctuations
for Markov processes with resetting (i.e., without any memory
of the pre-resetting history) have been derived~\cite{EM11b,EMM13,R16,HK16,PR17},
many fundamental questions concerning the
full first-passage probability under resetting, with or without memory effects,
still remain open~\cite{BMS13,BEM17}. 

Recent variations on the resetting theme have been to consider: resetting
of discrete-time L\'evy flights \cite{KMSS14} and continuous-time L\'evy walks 
\cite{KG15,CM15}, resetting for random walks in a bounded domain \cite{CS15,CM15},
resetting of extended systems such as fluctuating interfaces~\cite{GMS13} and
a reaction diffusion process in one dimension
\cite{DHP14}, Michaelis-Menten reaction schemes \cite{RRU15,R16},
the thermodynamics of resetting \cite{FGS16,PRahav17} and 
large deviations of the additive functionals of resetting processes 
\cite{MST15,HT17,HMMT18},
interaction-driven resetting \cite{FE17},
resetting with branching
\cite{PER18} and  fractional Brownian motion with resetting \cite{MO18}.
Very recently, resetting dynamics in quantum systems have also
been studied \cite{MSM18,RTLG18}.

In this work we employ a renewal equation approach first noted in \cite{EM11a} (see also 
\cite{KMSS14} for the computation of first-passage probability using the
renewal approach and  
\cite{CS18} for recent work). For the velocity 
randomization case  we may use a simple renewal equation
for the survival probability (\ref{Qrenew}), which is applicable to 
many systems resetting to their initial conditions. In the general velocity resetting case
a system of renewal equations for joint probabilities of survival and velocity is required (\ref{Qrenewgen}).

Our study reveals that resetting  of position and velocity of run and tumble particle
results in nonequilibrium stationary state that is a Laplace distribution
(symmetric, exponential decay) which is of the same form as a diffusive particle with position resetting.
However the survival probability of the path particle and the mean first passage time do depend on the velocity resetting parameter $\eta$.

The paper is organised as follows. In section 2 we review 
run and tumble dynamics as described by a Master equation system.
We then consider position resetting and velocity randomization and compute the 
stationary state in section 3, survival probability in section 4 and mean time to absorption in section 5. In section 6 we consider general velocity resetting parametrised by $\eta$ and present a general renewal scheme.
We work out particular formulae for the mean time to absorption for 
position-only resetting and compare to the velocity randomization case.
 We conclude in section 7.

\section{Run and tumble particle dynamics}
\label{sec:rnt}
In this section we review  the dynamics of 
a run and tumble particle (see for example \cite{Weiss,MJKKSMRD18,HV10}).
The system of forward master equations (in the absence of resetting) read
\begin{eqnarray}
\frac{\partial P_+(x,t)}{\partial t}
&=& -v_0 \frac{\partial P_+(x,t)}{\partial x} - \gamma P_+(x,t) + \gamma P_-(x,t) \label{Ppt} \\
\frac{\partial P_-(x,t)}{\partial t}
&=& +v_0 \frac{\partial P_-(x,t)}{\partial x} - \gamma P_-(x,t) + \gamma P_+(x,t)  \label{Pmt}
\end{eqnarray}
where $P_\sigma(x,t)$ is the probability density for the particle to
have velocity $\sigma$ and be at position $x$ at time $t$. The terms proportional to
$\gamma$ originate from  the switching of velocity with rate $\gamma$;
the correlation time of the velocity is thus $1/\gamma$.
We note that the system is invariant under time reversal:  $\sigma \to -\sigma$
and $v_0 \to - v_0$.

It will be convenient to have at our disposal the form of the Laplace transforms of $P_\pm$,
which are defined as
\begin{equation}
\tilde P_\pm(x, s) = \int_0^\infty \D t\, {\rm e}^{-st} P_\pm (x,t)\;.
\end{equation}
Taking the Laplace transform of (\ref{Ppt},\ref{Pmt}) we obtain the system
\begin{eqnarray}
-P_+(x,0)
+ v_0 \frac{\D \tilde P_+}{\D x} +(s+ \gamma) \tilde P_+ - 
\gamma \tilde P_- &=&0  \label{Pps} \\
-P_-(x,0)
- v_0 \frac{\partial \tilde P_+}{\partial x} +(s+ \gamma) \tilde P_- - 
\gamma \tilde P_+ &=&0  \label{Pms} \;.
\end{eqnarray}

We  need to fix initial conditions which we choose to be at the origin and symmetric
\begin{equation}
P_+(x,0) = P_-(x,0) = \frac{1}{2} \delta(x)\;,
\end{equation}
i.e. the particle begins at the origin with equal probability for the  velocity
$\sigma(0) = \pm 1$.

By taking a further spatial derivative and rearranging,
we may turn  the first-order system (\ref{Pps}, \ref{Pms}) 
into  {\em decoupled} second-order equations  
which read (away from $x=0$)
\begin{equation}
v_0^2 \frac{\D^2 \tilde P_\pm}{\D x^2}
- \left[ (s+\gamma)^2 -\gamma^2\right] \tilde P_{\pm} =0\;.
\end{equation}

The solutions which respect the boundary conditions that $P_{\pm}$
remain finite as $x \to \pm \infty$ are
\begin{eqnarray}
\tilde P_\pm= A_\pm {\rm e^{-\lambda x}}  \quad\mbox{for}\quad x>0 \label{coeffA}\\
\tilde P_\pm = B_\pm {\rm e^{ +\lambda x}}  \quad\mbox{for}\quad x<0 \label{coeffB}
\end{eqnarray}
where
\begin{equation}
\lambda = \left( \frac{s(s+2\gamma)}{v_0^2}\right)^{1/2}\;.
\label{lamdef}
\end{equation}
In order to fix the coefficients $A_\pm$, $B_\pm$ we go back to 
(\ref{Pps}, \ref{Pms}) and obtain conditions
\begin{eqnarray}
(s+\gamma -\lambda v_0) A_+ =\gamma A_- \label{Acond} \\
(s+\gamma +\lambda v_0) B_+ =\gamma B_- \label{Bcond} \;.
\end{eqnarray}
Also, as the initial condition is symmetric around $x=0$ and the dynamics is invariant under time reversal, the total probability $P(x,t) = P_+(x,t) + P_-(x,t)$
must be symmetric about $x=0$. This implies
\begin{eqnarray}
A_+ + A_- = B_+ + B_-\;.\label{AB1}
\end{eqnarray}
Finally, normalisation of probability dictates
\begin{equation}
\int \D x \left[ \tilde P_+ + \tilde P_-\right] = \frac{1}{s}
\end{equation}
which implies
\begin{equation}
A_+ + A_- = B_+ + B_- = \frac{\lambda}{2s}\; \label{AB2}
\end{equation}
so that
\begin{equation}
\tilde P(x,s) = \tilde P_+(x,s)+ \tilde P_-(x,s)
= \frac{\lambda }{2s} {\rm e}^{-\lambda |x|}\;,
\label{Plt}
\end{equation}
where $\lambda=\sqrt{s(s+2\gamma)}/v_0$.
The Laplace transform (\ref{Plt}) is sufficient for our purposes in the 
next section. 
However, it is possible to invert the Laplace transform 
\cite{Othmer,Martens,Weiss} to obtain the time-dependent distribution 
\begin{equation}
P(x,t) =\frac{{\rm e}^{-\gamma t}}{2} \left\{ \delta(x-v_0 t)
+\delta(x+v_0 t) 
+ \frac{\gamma}{v_0}\left[ I_0(\rho) + \frac{\gamma\,t\, 
I_1(\rho)}{\rho}\right] \Theta(v_0t-|x|)
\right\}
\end{equation}
where
\begin{equation}
\rho = \sqrt{v_0^2 t^2 - x^2}\,\frac{\gamma}{v_0}
\end{equation}
and $I_0(\rho)$ and $I_1(\rho)$ are modified Bessel functions of the first kind.
(Note that in Ref. \cite{Weiss} there are some misprints.) To derive this
result, one can formally invert the Laplace 
transform in Eq. (\ref{Plt}) and write it as a Bromwich integral in the 
complex $s$ plane
\begin{equation}
P(x,t)= \int_{\Gamma} \frac{ds}{2\pi i}\, e^{st}\, \frac{\lambda(s)}{2s}\, 
e^{-\lambda(s)\, |x|}
\label{brom.1}
\end{equation}
where $\lambda(s)= \sqrt{s(s+2\gamma)}/v_0$ is a function of $s$ and the
Bromwich contour $\Gamma$ is a vertical line with 
its real part to the
right of all singularities of the integrand. The integrand has a branch 
cut over $s\in [-2\gamma,0]$ along the negative real $s$ axis.
One can close the contour in the left half plane and evaluate
the branch cut integral. Lengthy algebra and the
use of integral representation of the modified Bessel function $I_0$, finally 
leads to the result in Eq. (\ref{brom.1}). Since it is a bit peripheral
to our interest here and the result is well known in the literature, we do 
not give the details of this computation.
 
Equation (\ref{Plt}) is the main result of this section. For completeness
we present the coefficients appearing in (\ref{coeffA}, \ref{coeffB})
which can be found by solving (\ref{Acond}, \ref{Bcond}, \ref{AB2})
\begin{eqnarray}
A_+ = \frac{\lambda \gamma}{2s(s+ 2 \gamma -\lambda v_0)}
&\quad& A_- = \frac{\lambda(s+  \gamma -\lambda v_0)}{2s(s+ 2 \gamma -\lambda v_0)}\\
B_+ = \frac{\lambda \gamma}{2s(s+ 2 \gamma +\lambda v_0)}
&\quad& B_- = \frac{\lambda(s+  \gamma +\lambda v_0)}{2s(s+ 2 \gamma +\lambda v_0)}\;.
\end{eqnarray}

\section{Run and tumble particle under position resetting and velocity randomization}
\label{sec:stst}
We now add a resetting process to the dynamics, which comprises simultaneous  resetting both the position and the velocity. In this section the resetting position is taken to be the origin.
With rate $r$ the particle resets its initial position and the velocity $\sigma$ is chosen to be $\pm 1$ with probability $1/2$, i.e. the velocity is 
randomized. We refer to this resetting protocol as position resetting and velocity randomization. We shall consider more general resetting protocols in section \ref{sec:genreset}.

Given that the initial condition of the particle is also at the origin with
the velocity $\sigma$ is chosen to be $\pm 1$ with probability $1/2$, we may write down a renewal equation \cite{EM11a} for the total probability density in the presence of resetting
which we denote $P_r(x,t)$ :
\begin{eqnarray}
P_r(x,t) &=& {\rm e}^{-rt} P_0(x,t) + r \int_0^t\D \tau {\rm e}^{-r\tau} P_0(x,\tau)\int \D x' P_r(x',t-\tau)  \label{renew1} \\
&=& {\rm e}^{-rt} P_0(x,t) + r \int_0^t\D \tau {\rm e}^{-r\tau} P_0(x,\tau)\;.
\label{renew2}
\end{eqnarray}
Here, $P_0(x,t)$ is the probability density  without resetting considered in section \ref{sec:rnt}.
The first term on the r.h.s of (\ref{renew1}) is the contribution from trajectories in which there is no resetting, which occurs with probability ${\rm e}^{-rt}$; the second term integrates the
contributions from trajectories in which the last reset occurs at time
$t-\tau$ and at position $x'$ and there is no resetting from time $t-\tau$ to $t$, which occurs
with probability ${\rm e}^{-r\tau}$. The second equality (\ref{renew2}) comes from the fact that when there are no absorbing boundaries  probability is conserved
which implies $\int \D x' P_r(x',t-\tau) =1$.
\subsection{Stationary State and limits}
The stationary distribution
\begin{equation}
P^{st}_r(x) = \lim_{t\to \infty} P_r(x,t) 
\end{equation}
is easily obtained from the limit $t\to \infty$ of (\ref{renew2}) from which we learn, using the result (\ref{Plt}) of section \ref{sec:rnt}, that
\begin{equation}
P^{st}_r(x) = r \tilde P(x,r) = \frac{\lambda_r}{2} {\rm e}^{-\lambda_r |x|}\;,
\label{Pstst}
\end{equation}
where $\lambda_r$ is given by
\begin{equation}
\lambda_r = \left( \frac{r(r+2\gamma)}{v_0^2}\right)^{1/2}\;.
\label{lamrdef}
\end{equation}

The distribution is a double  exponential distribution, also known as Laplace distribution, with decay length
\begin{equation}
\ell \equiv \frac{1}{\lambda_r} = 
\left( \frac{v_0^2}{r(r+2\gamma)}\right)^{1/2}\;.
\label{decay}
\end{equation}
Thus the decay length increases with the speed $v_0$ but decreases with switching rate $\gamma$ and resetting rate $r$. 

It is of interest to consider the various limits of the process
and the form of the decay length in these limits.
First, in the limit of no resetting  $r\to 0$, the decay length diverges as $r^{-1/2}$ indicating that there is no stationary state.
The {\em ballistic limit} is when the switching rate $\gamma \to 0$ in which case
$\ell \to \frac{v_0}{r}$ which is the mean distance travelled between resets. Finally the {\em diffusive limit} occurs when both $v_0$ and $\gamma$
diverge but
\begin{equation}
\lim_{v_0, \gamma \to \infty} \frac{v_0^2}{\gamma}  = 2D
\label{difflim}
\end{equation}
where $D$ is the diffusion coefficient. Then
$\ell \to \sqrt{\frac{D}{r}}$ which recovers the expression for
diffusive resetting \cite{EM11a}.

\section{Survival Probability} 
We now consider the survival probability of the persistent random walker
in the presence of an absorbing boundary at the origin
(and under position resetting to  $X_r\ge 0$ 
and velocity randomization as in section~\ref{sec:stst}).
In the context of a search we refer to the origin as the target; clearly, the event of particle touching the boundary corresponds to the event of a searcher locating the target. 

We again  take advantage of a renewal equation.
We first define
$Q_r (x_0,t)$  as the survival probability in the {\em presence} of resetting
and $Q_0 (x_0,t)$ as  the survival probability in the {\em absence} of resetting,  for
a particle having started from initial position $x=x_0\ge 0$ with initial velocity 
chosen to be $\sigma = \pm 1$ with equal probability $1/2$. Note that
$Q_r (x_0,t)$  
implies an integration over the final position of the particle. 
Also note that the initial position $x_0$ is a variable which, 
at the end of the calculation, we may set equal to $X_r$.

Then we have a renewal equation analogous to
(\ref{renew1})
\begin{equation}
Q_r(x_0,t) = {\rm e}^{-rt} Q_0(x_0,t) + r \int_0^t\D \tau {\rm e}^{-r\tau} Q_0(X_r,\tau) Q_r(x_0, t-\tau)\;.
\label{Qrenew}
\end{equation}
Again, the first term on the r.h.s is the contribution from survival trajectories in which there is no resetting; the second term integrates the
contributions from survival trajectories in which the last reset occurs at time
$t-\tau$.

Taking the Laplace Transform
\begin{equation}
\tilde  Q_*(x_0, s) = \int_0^\infty \D t\, {\rm e}^{-st} Q_*(x_0,t)\;,
\end{equation}
where $*$ indicates  $0$ or $r$, we readily obtain from (\ref{Qrenew})
\begin{equation}
\tilde Q_r(x_0,s) = 
\frac{ \tilde Q_0(x_0,r+s)}{ 1- r \tilde Q_0(X_r,r+s)} 
\end{equation}
and, in particular, setting the initial position $x_0 = X_r$,
\begin{equation}
\tilde Q_r(X_r,s) = 
- \frac{1}{r} +\frac{ 1 }{ r(1- r \tilde Q_0(X_r,r+s))} \;.
\label{Qr}
\end{equation}
Equation (\ref{Qr}) is an equation of rather general applicability, which applies
whenever resetting to the initial conditions is a  Poisson process with rate $r$.
 
\subsection{Survival probability in the absence of resetting}
\label{sec:survnores}
In view of (\ref{Qr}) we just need to compute $\tilde Q_0(X_r,s)$, the Laplace transform
of  the survival probability in the absence of resetting. 
This was computed recently in Ref. \cite{MJKKSMRD18}. We reproduce
it here for the sake of completeness.
Following \cite{MJKKSMRD18}, we introduce  
$Q_0^{+} (x_0,t)$ and $Q_0^{-} (x_0,t)$ as the  survival probability (without 
resetting)
for a particle having started from position $x=x_0\ge 0$ with  initial velocity
$\pm 1$ respectively.

We can write down a system of {\em backward} evolution equations
for these survival probabilities
\begin{eqnarray}
\frac{\partial Q_0^+(x_0,t)}{\partial t} = v_0
\frac{\partial Q_0^+(x_0,t)}{\partial x_0}  - \gamma  Q_0^+(x_0,t) + \gamma Q_0^-(x_0,t) \label{Qpt} \\
\frac{\partial Q_0^-(x_0,t)}{\partial t} = -v_0
\frac{\partial Q_0^-(x_0,t)}{\partial x_0}  - \gamma Q_0^-(x_0,t) + \gamma Q_0^+(x_0,t)\;.  \label{Qmt} 
\end{eqnarray}
which needs to be solved in the positive half-space $x_0\ge 0$.
The initial conditions are $Q_0^+(x_0,0)= Q_0^-(x_0,0)=1$
and the boundary condition, which imposes an absorbing boundary at $x=0$, is just
$Q_0^-(0,t)=0$. This is because if the particle starts at the origin with a
negative initial velocity it can not survive up to finite time $t$. In contrast,
if it starts with a positive velocity, it can survive and $Q_0^+(0,t)$ is therefore
unspecified and has to be determined a posteriori. In fact, as we will see below
that just the single condition $Q_0^{-}(0,t)=0$ is sufficient to provide
a unique solution to this system of coupled equations.

Taking the Laplace transform of (\ref{Qpt},\ref{Qmt}) yields
\begin{eqnarray}
+ v_0 \frac{\partial \tilde Q_0^+}{\partial x_0} -( s+ \gamma) \tilde Q_0^+ +
\gamma \tilde Q_0^-  = -1\\
- v_0 \frac{\partial \tilde Q_0^-}{\partial x_0} -( s+ \gamma) \tilde Q_0^- +
\gamma \tilde Q_0^+  = -1 \label{Qmlt}
\end{eqnarray}
from which a further spatial derivative and rearrangement yields
the decoupled equation
\begin{equation}
 v_0^2 \frac{\partial^2 \tilde Q_0^-}{\partial x_0^2} -s( s+ 2\gamma) \tilde Q_0^-  = 
-\frac{(2\gamma +s)}{v_0}\;.
\end{equation}
The solution which satisfies the boundary condition $Q_-(0,t)=0$ is
\begin{equation}
\tilde Q_0^-(x_0, s) = \frac{1}{s}\left[ 1- {\rm e}^{-\lambda x_0}\right]
\end{equation}
where $\lambda$ is given by (\ref{lamdef}).
Substituting back into (\ref{Qmlt}) yields 
\begin{equation}
\tilde Q_0^+(x_0, s) = \frac{1}{s} + \frac{1}{\gamma s}
\left[v_0 \lambda - (s+\gamma)\right] {\rm e}^{-\lambda x_0}\;.
\end{equation}
Given the symmetric velocity initial condition, we  have
\begin{eqnarray}
\tilde Q_0(x_0, s) &\equiv& \frac{1}{2}\left[ \tilde Q_0^+(x_0, s)+ \tilde Q_0^-(x_0, s)\right]\nonumber \\
& =&
 \frac{1}{s} + \frac{1}{2\gamma s}
\left[v_0 \lambda - (s+2\gamma)\right] {\rm e}^{-\lambda x_0}\;.
\label{Qlt}
\end{eqnarray}
Inserting (\ref{Qlt}) into (\ref{Qr}) yields the result
\begin{equation}
\tilde Q_r(X_r, s) 
= -\frac{1}{r} + \frac{1}{r} \left[
\frac{2\gamma (s+r) {\rm e}^{\lambda_{s+r} X_r}}{
2\gamma s {\rm e}^{\lambda_{s+r} X_r} - r \left[v_0 \lambda_{r+s} - (r+s+2\gamma)\right]} \right]
\label{Qlf}
\end{equation}
where
\begin{equation}
\lambda_{r+s} = \left( \frac{(r+s)(r+s+2\gamma)}{v_0^2}\right)^{1/2}\;.
\label{lamrsdef}
\end{equation}

\section{Mean first passage time}
The mean first passage time to the origin (or equivalently the mean time to absorption at the origin), $T(X_r)$, is conveniently given by
\begin{equation}
T(X_r) = \tilde Q(X_r, s=0)\;.
\end{equation}
In the $s\to 0$ limit it can be checked that (\ref{Qlf}) reduces to
\begin{eqnarray}
T(X_r) =  -\frac{1}{r}
+\frac{2\gamma}{r} \left[\frac{{\rm e}^{\lambda_r X_r}}{r+2\gamma -(r(r+2\gamma))^{1/2}}
\right]
 \end{eqnarray}
where $\lambda_r$ is given by (\ref{lamrdef}).

First let us check the diffusive limit (\ref{difflim})
in which case $\lambda_r \to  (r/D)^{1/2}$
and
\begin{equation}
T(X_r) \to  - \frac{1}{r} + \frac{{\rm e}^{(r/D)^{1/2}X_r}}{r} 
\label{diff_limit}
\end{equation}
recovering the result of \cite{EM11a}.

We also note that $T(X_r)$ diverges as $r^{-1/2}$ as $r\to 0$
and also diverges exponentially in $r$ as $r\to \infty$ implying a minimum value at intermediate $r$. In order to analyse where this minimum occurs it is useful to introduce reduced  variables
\begin{eqnarray} 
R&=&  \frac{r}{2 \gamma} \label{Rdef}\\
\xi &=& \frac{2 \gamma X_r}{v_0} \label{xidef}\;.
\end{eqnarray}
$R$ is half the ratio of resetting rate to velocity switching rate
whereas $\xi$ is twice the ratio of distance to the target to  mean distance travelled  between reversals of velocity (the factors of two are included for later convenience).
In terms of these variables
\begin{eqnarray}
\lambda_r X_r &=& (R(R+1))^{1/2}\xi
\end{eqnarray}
and one obtains
\begin{eqnarray}
2\gamma T(R,\xi) &=&  -\frac{1}{R} 
+ \frac{ {\rm e}^{(R(1+R))^{1/2}\xi}}{ R\left[1+R -(R(1+R))^{1/2} \right]}\;.
\label{Tred}
\end{eqnarray}
We may minimise this expression with respect to $R$ at fixed $\xi$. It has a
unique minimum. A plot of $T(R,\xi=1)$ vs. $R$ is shown in Fig. (\ref{fig.TR1}).

\section{General velocity resetting}
\label{sec:genreset}
We now consider a more general resetting process  which comprises simultaneous  resetting of both the position and the velocity.
With rate $r$ the particle resets its initial position at $X_r$ and the velocity $\sigma$ is reversed to $-\sigma$ with probability $\eta$ or remains $\sigma$ with probability $1-\eta$. The case $\eta =1/2$ corresponds to the velocity randomization
considered in earlier sections and the case $\eta =0$ corresponds to position-only resetting.

The first thing to note is that given a {\em symmetric} initial condition the stationary state is independent of $\eta$. The reason is that the velocity distribution will remain symmetric and is independent of $\eta$.
To demonstrate this explicitly
we let $P_*^{\sigma_f \sigma_i}(x,t)$ be the probability density of being at $x$ at time $t$ and having velocity $\sigma_f$ given that the particle began at $t=0$ at $X_r$ with velocity $\sigma_i$;
$*$ indicates  $0$ or $r$ and corresponds to no resetting or with resetting respectively. We may then write down the following renewal equation system
\begin{eqnarray}
P_r^{\sigma_f\, \sigma_i}(x,t) &=& {\rm e}^{-rt} P_0^{\sigma_f\, \sigma_i}(x,t)
\label{Prenewgen} \\
&+&  r \int_0^t \D \tau {\rm e}^{-r\tau}\int \D x' \left\{
 P_r^{\sigma_f\, \sigma_i}(x', t-\tau) \left[ (1-\eta) P_0^{\sigma_f\, \sigma_f}(x,\tau)
+ \eta  P_0^{\sigma_f\, -\sigma_f}(x, \tau)\right]\nonumber \right. \\
&&
 \left. 
\hspace*{2.2cm}+  P_r^{-\sigma_f\, \sigma_i}(x', t-\tau)\left[ (1-\eta) P_0^{\sigma_f\, -\sigma_f}(x, \tau)
+ \eta  P_0^{\sigma_f\, \sigma_f}(x, \tau)\right]\right\} \nonumber\;.
\end{eqnarray}
Now let us fix 
the initial conditions at $t=0$ as  $\sigma = \pm 1 $ with probability $1/2$
and define
\begin{eqnarray}
P_*^{\sigma_f}(x, t) = \frac{1}{2} P_*^{\sigma_f\, +}(x, t) + \frac{1}{2} P_*^{\sigma_f\, -}(x,t)\;.
\end{eqnarray}
The system (\ref{Prenewgen}) becomes
\begin{eqnarray}
P_r^{\sigma_f}(x,t) &=& {\rm e}^{-rt} P_0^{\sigma_f}(x,t)
\label{Prenewgen2} \\
&& + r \int_0^t \D \tau {\rm e}^{-r\tau}\int \D x' \left\{
P_r^{\sigma_f}(x', t-\tau) \left[ (1-\eta) P_0^{\sigma_f\, \sigma_f}(x,\tau)
+ \eta  P_0^{\sigma_f\, -\sigma_f}(x, \tau)\right]\nonumber \right. \\
&&
 \left. 
\hspace*{2.2cm}+  P_r^{-\sigma_f}(x', t-\tau)\left[ (1-\eta) P_0^{\sigma_f\, -\sigma_f}(x, \tau)
+ \eta  P_0^{\sigma_f\, \sigma_f}(x, \tau)\right]\right\} \nonumber\;.
\end{eqnarray}
Now  due to the symmetric initial condition we have
$\int \D x' P_r^{\sigma_f}(x, t) =1/2$
and we find that the terms with coefficient $\eta$ in (\ref{Prenewgen2}) cancel,
leaving
\begin{eqnarray}
P_r^{\sigma_f}(x,t) &=& {\rm e}^{-rt} P_0^{\sigma_f}(x,t) \\
&& + \frac{r}{2} \int_0^t \D \tau {\rm e}^{-r\tau}\left\{
P_0^{\sigma_f\, \sigma_f}(x,\tau)
+  P_0^{\sigma_f\, -\sigma_f}(x, \tau) \right\}
\end{eqnarray}
which recovers (\ref{renew2}). Thus, the stationary state of the resetting run and tumble particle does not depend on the velocity resetting protocol.

However, as we shall now show the survival probability in the presence of an absorbing boundary does depend on $\eta$.

\subsection{Survival probability}

In order to  solve for  the survival probability in the general case we need to extend previous survival probability results to the computation of joint survival and final velocity distributions. We define
$Q_r^{\sigma_f \sigma_i}(t)$ and $Q_0^{\sigma_f \sigma_i}(t)$
as the joint probability of survival and having velocity $\sigma_f$ at time $t$, given that the particle began at $X_r$ with velocity $\sigma_i$, with and without resetting respectively.  To ease the notation we shall drop the dependence on initial position
(which is always $X_r$) from the $Q_r^{\sigma_f \sigma_i}(t)$.

Then we may write down a renewal equation system as follows
\begin{eqnarray}
Q_r^{\sigma_f\, \sigma_i}(t) &=& {\rm e}^{-rt} Q_0^{\sigma_f\, \sigma_i}(t)
\label{Qrenewgen} \\
&& + r \int_0^t \D \tau {\rm e}^{-r\tau}\left\{
Q_r^{\sigma_f\, \sigma_i}(t-\tau) \left[ (1-\eta) Q_0^{\sigma_f\, \sigma_f}(\tau)
+ \eta  Q_0^{\sigma_f\, -\sigma_f}(\tau)\right]\nonumber \right. \\
&&
 \left. 
\hspace*{2.2cm}+ Q_r^{-\sigma_f\, \sigma_i}(t-\tau)\left[ (1-\eta) Q_0^{\sigma_f\, -\sigma_f}(\tau)
+ \eta  Q_0^{\sigma_f\, \sigma_f}(\tau)\right]\right\} \nonumber\;.
\end{eqnarray}
Again this equation is easily understood: the first term represents surviving trajectories within which no resetting occurred; the second term integrates up the surviving trajectories which have the last reset at time $t-\tau$ and the coefficients $(1-\eta)$ and $\eta$ give the probability of a velocity switch occurring at that reset.

We now take the Laplace transform
\begin{equation}
\tilde  Q_*^{\sigma_f\, \sigma_i}(s) = \int_0^\infty \D t\, {\rm e}^{-st} Q_*^{\sigma_f\, \sigma_i}(t)
\end{equation}
with $* = 0,r$,
to obtain
\begin{eqnarray}
\tilde Q_r^{\sigma_f\, \sigma_i}(s) &=& \tilde Q_0^{\sigma_f\, \sigma_i}(r+s)\label{Qltgen} \\
&& + r \left\{
Q_r^{\sigma_f\, \sigma_i}(s) \left[ (1-\eta) \tilde Q_0^{\sigma_f\, \sigma_f}(r+s)
+ \eta  \tilde Q_0^{\sigma_f\, -\sigma_f}(r+s)\right]\nonumber \right. \\
&&
 \left. 
\hspace*{2.2cm}+ \tilde Q_r^{-\sigma_f\, \sigma_i}(s)\left[ (1-\eta) \tilde Q_0^{\sigma_f\, -\sigma_f}(r+s)
+ \eta  \tilde Q_0^{\sigma_f\, \sigma_f}(r+s)\right]\right\} \nonumber
\end{eqnarray}

As usual 
the initial conditions at $t=0$ are  $\sigma = \pm 1 $ with probability $1/2$
and we define
\begin{eqnarray}
Q_*^{\sigma_f}(t) = \frac{1}{2} Q_*^{\sigma_f\, +}(t) + \frac{1}{2} Q_*^{\sigma_f\, -}(t)
\end{eqnarray}
with similar definitions for the Laplace transforms. Then system
(\ref{Qltgen}) becomes (where we now write out explicitly the two equations) 
\begin{eqnarray}
\tilde Q_r^{+}(s) &=& \tilde Q_0^{+}(r+s) \label{Qltgenp}\\
&& + r \left\{
Q_r^{+}(s) \left[ (1-\eta) \tilde Q_0^{+\, +}(r+s)
+ \eta  \tilde Q_0^{+\, -}(r+s)\right]\nonumber \right.  \\
&&
 \left. 
\hspace*{2.2cm}+ \tilde Q_r^{-}(s)\left[ (1-\eta) \tilde Q_0^{+\, -}(r+s)
+ \eta  \tilde Q_0^{+\, +}(r+s)\right]\right\} \nonumber \;,\\
\tilde Q_r^{-}(s) &=& \tilde Q_0^{-}(r+s)  \\
&& + r \left\{
Q_r^{-}(s) \left[ (1-\eta) \tilde Q_0^{-\, -}(r+s)
+ \eta  \tilde Q_0^{-\, +}(r+s)\right]\nonumber \right. \\
&&
 \left. 
\hspace*{2.2cm}+ \tilde Q_r^{+}(s)\left[ (1-\eta )\tilde Q_0^{-\, +}(r+s)
+ \eta  \tilde Q_0^{-\, -}(r+s)\right]\right\} \nonumber
\end{eqnarray}
This system is easily solved to give
\begin{eqnarray}
\tilde Q_r^+(s) = \frac{1}{ad-bc} \left[ d\tilde Q_0^+(r+s)- b\tilde Q_0^-(r+s)\right]\;.
\label{Qltgenpsol}\\
\tilde Q_r^-(s) = \frac{1}{ad-bc} \left[ -c\tilde Q_0^+(r+s) +a\tilde Q_0^-(r+s)\right]
\label{Qltgenmsol}
\end{eqnarray}
where
\begin{eqnarray}
a &=& 1- r\left[(1-\eta)\tilde Q_0^{+\,+}(r+s)+ \eta\tilde  Q_0^{+\,-}(r+s)\right] \\
b &=& - r\left[ \eta \tilde Q_0^{+\, +}(r+s) + (1-\eta) \tilde Q_0^{+\, -}(r+s)\right]\\
c &=& - r\left[ \eta \tilde Q_0^{-\, -}(r+s) + (1-\eta) \tilde Q_0^{-\, +}(r+s)\right]\\
d  &=& 1- r\left[(1-\eta)\tilde Q_0^{-\,-}(r+s)+ \eta\tilde  Q_0^{-\,+}(r+s)\right] \;.
\end{eqnarray}

Thus we obtain the general expression for the Laplace transform of the  total survival probability 
\begin{eqnarray}
\tilde Q_r(s) &\equiv& \tilde Q^+_r(s) + \tilde Q^-_r(s) \\
&=& \frac{1}{ad-bc} \left[ (d-c)\tilde Q_0^+(r+s)+(a- b)\tilde Q_0^-(r+s)\right]\;.
\label{Qltgensol}
\end{eqnarray}

The solution (\ref{Qltgensol}) simplifies greatly
when $\eta=1/2$ in which case $d-c =1$, $a-b=1$
and $ad-bc = 1-r[Q_0^+(r+s) + Q_0^-(r+s)] = 1-rQ_0(r+s) $
and the result (\ref{Qr}) is recovered.

In the case of  general $\eta$ we require the knowledge of the Laplace transforms
$\tilde Q_0^{\sigma_f \sigma_i}$ which we now show how to compute.

\subsection{Survival probabilities in absence of reset}
We  generalise the system (\ref{Qpt}, \ref{Qmt}) of section~\ref{sec:survnores}
we write down a system of four backward equations as
\begin{equation}
\frac{\partial Q_0^{\sigma_f\, \sigma_i}(x_0,t)}{\partial t} = 
\sigma_i v_0 \frac{\partial Q_0^{\sigma_f\, \sigma_i}(x_0, t)}{\partial x_0}  - \gamma  Q_0^{\sigma_f\, \sigma_i}(x_0, t) + \gamma Q_0^{\sigma_f\, -\sigma_i}(x_0, t) \label{Qgent} 
\label{bfp4}
\end{equation}
Note that we have kept here the explicit $x_0$ dependence in $Q_0^{\sigma_f 
\sigma_i}(x_0,t)$ since we use $x_0$ as a variable in the backward Fokker-Planck
approach. Eq. (\ref{bfp4}) has to be solved in the domain $x_0\ge 0$.
The initial conditions are now
\begin{eqnarray}
Q_0^{++}(x_0, 0) = Q_0^{--}(x_0, 0) = 1\\
Q_0^{+-}(x_0, 0) = Q_0^{-+}(x_0, 0) =  0
\end{eqnarray}
and the boundary condition corresponding to the absorbing boundary at $x_0=0$ is
\begin{equation}
Q_0^{+-}(0, t) = Q_0^{--}(0, t) =0\;.
\label{Qssbc}
\end{equation}
As usual, the solution to the system (\ref{Qgent}) is obtained by Laplace transform
which we write out explicitly to show that it breaks into two subsystems
\begin{eqnarray}
-1 
&=& v_0 \frac{\partial \tilde Q_0^{+\,+}(x_0,s)}{\partial x_0} - (s+\gamma) \tilde 
Q_0^{+\,+}(x_0,s) + \gamma \tilde Q_0^{+\,-}(x_0,s) \label{Qgenltpp} \\
0
&=& -v_0 \frac{\partial \tilde Q_0^{+\,-}(x_0,s)}{\partial x_0} - (s+\gamma) \tilde 
Q_0^{+\,-}(x_0,s) + \gamma \tilde Q_0^{+\,+}(x_0,s) \label{Qgenltpm}\\
-1
&=& -v_0 \frac{\partial \tilde Q_0^{-\,-}(x_0,s)}{\partial x_0} - (s+ \gamma) \tilde 
Q_0^{-\,-}(x_0,s) + \gamma \tilde Q_0^{-\,+}(x_0,s) \label{Qgenltmp}\\
0 &=& v_0 \frac{\partial \tilde Q^{-\,+}(x_0,s)}{\partial x_0} - (s +\gamma) \tilde 
Q_0^{-\,+}(x_0,s) + \gamma \tilde Q_0^{-\,-}(x_0,s) \label{Qgenltmm}
\end{eqnarray}
Then equations (\ref{Qgenltpp}) and (\ref{Qgenltpm})
and equations (\ref{Qgenltmp}) and (\ref{Qgenltmm})  can be turned into decoupled second order equations
\begin{eqnarray}
\frac{\partial^2 \tilde Q_0^{+-}(x_0,s)}{\partial x_0^2}
&=& \lambda^2 \tilde Q_0^{+-}(x_0,s) -\frac{\gamma}{v_0^2}\\
\frac{\partial^2 \tilde Q_0^{--}(x_0,s)}{\partial x_0^2}
&=& \lambda^2 \tilde Q_0^{--}(x_0,s) -\frac{s+\gamma}{v_0^2}
\end{eqnarray}
The solution satisfying the boundary condition (\ref{Qssbc}) is
\begin{eqnarray}
\tilde Q_0^{+\, -}(s) &=& \frac{\gamma}{s(s+2\gamma)}\left[1- {\rm e}^{-\lambda 
x_0}\right]\label{Q0pm}\\
\tilde Q_0^{+\, +}(s) &=& \frac{s+\gamma}{s(s+2\gamma)}\left[1- {\rm e}^{-\lambda 
x_0}\right]   +  \frac{v_0\lambda}{s(s+2\gamma)}{\rm e}^{-\lambda x_0} \\
\tilde Q_0^{-\, -}(s) &=& \frac{s+\gamma}{s(s+2\gamma)}\left[1- {\rm e}^{-\lambda x_0}\right]\\
\tilde Q_0^{-\, +}(s) &=& -\frac{1}{\gamma} + \frac{(s+\gamma)^2}{\gamma s(s+2\gamma)}\left[1- {\rm e}^{-\lambda x_0}\right]  +    \frac{v_0(s+\gamma)\lambda}{ \gamma s(s+2\gamma)}{\rm e}^{-\lambda x_0}
\label{Q0mp} \;.
\end{eqnarray}
where we have dropped, as usual for brevity, the explicit $x_0$ 
dependence of $\tilde Q_0^{\sigma_f \sigma_i}(x_0,s)\equiv \tilde Q_0^{\sigma_f 
\sigma_i}(s)$.

\subsection{Position-only resetting}
As a specific example we present results for the case $\eta=0$ i.e. position-only resetting.

After unilluminating algebra (which we do not present here)
equation (\ref{Qltgensol}) may be reduced to:
\begin{eqnarray}
\tilde Q_r(s) &=&
-\frac{1}{r} + \frac{1}{r} \frac{1-\frac{r}{2}\left[\tilde Q_0^{++}(r+s)+\tilde Q_0^{--}(r+s)
-\tilde Q_0^{+-}(r+s)-\tilde Q_0^{-+}(r+s)\right]}
{ad-bc}\nonumber\\
\\
ad-bc &=&
1-r\left[\tilde Q_0^{++}(r+s)+\tilde  Q_0^{--}(r+s)\right]\nonumber\\
&&+r^2\left[\tilde Q_0^{++}(r+s) \tilde Q_0^{--}(r+s) -\tilde  Q_0^{+-}(r+s)\tilde  Q_0^{-+}(r+s)\right]
\end{eqnarray}
Using expressions (\ref{Q0pm}--\ref{Q0mp}) one eventually obtains
\begin{equation}
\tilde Q(s) = -\frac{1}{r}+
\frac{1}{r^2E}\left[ \frac{s+2\gamma}{r}  +\frac{(1-\beta)}{2}{\rm e}^{-\lambda_{r+s} X_r}\right]  \label{Qtotres}
\end{equation}
where
\begin{eqnarray}
\beta &=& \frac{\gamma}{\lambda_{r+s} v_0 + \gamma + r +s}
\end{eqnarray}
and
\begin{eqnarray}
E &=&  \frac{(s+2\gamma)s}{r(r+s)}
+\frac{\left[ (\beta+1) \gamma+ s \right]}{r(r+s)}\, {\rm e}^{-\lambda_{r+s} X_r}\label{Edef}
\end{eqnarray}
where $\lambda_{r+s}$ is given in Eq. (\ref{lamrsdef}).
The mean first passage time to the origin (or equivalently the mean time to absorption at the origin), $T(X_r)$, is conveniently given by
the $s\to 0$ limit of (\ref{Qtotres}) which reduces to
\begin{eqnarray}
T(X_r) = -\frac{1}{r}+ \frac{1}{r+2\gamma -(r(r+2\gamma))^{1/2}}
\left[ \frac{2\gamma  {\rm e}^{\lambda X_r} }{r}
-\frac{r}{2\gamma} + \frac{(r(r+2\gamma))^{1/2}}{2\gamma}\right] \nonumber\\
 \end{eqnarray}
where $\lambda$ is given by (\ref{lamdef}).

\begin{figure}[t]
\centering
\includegraphics[scale=0.5]{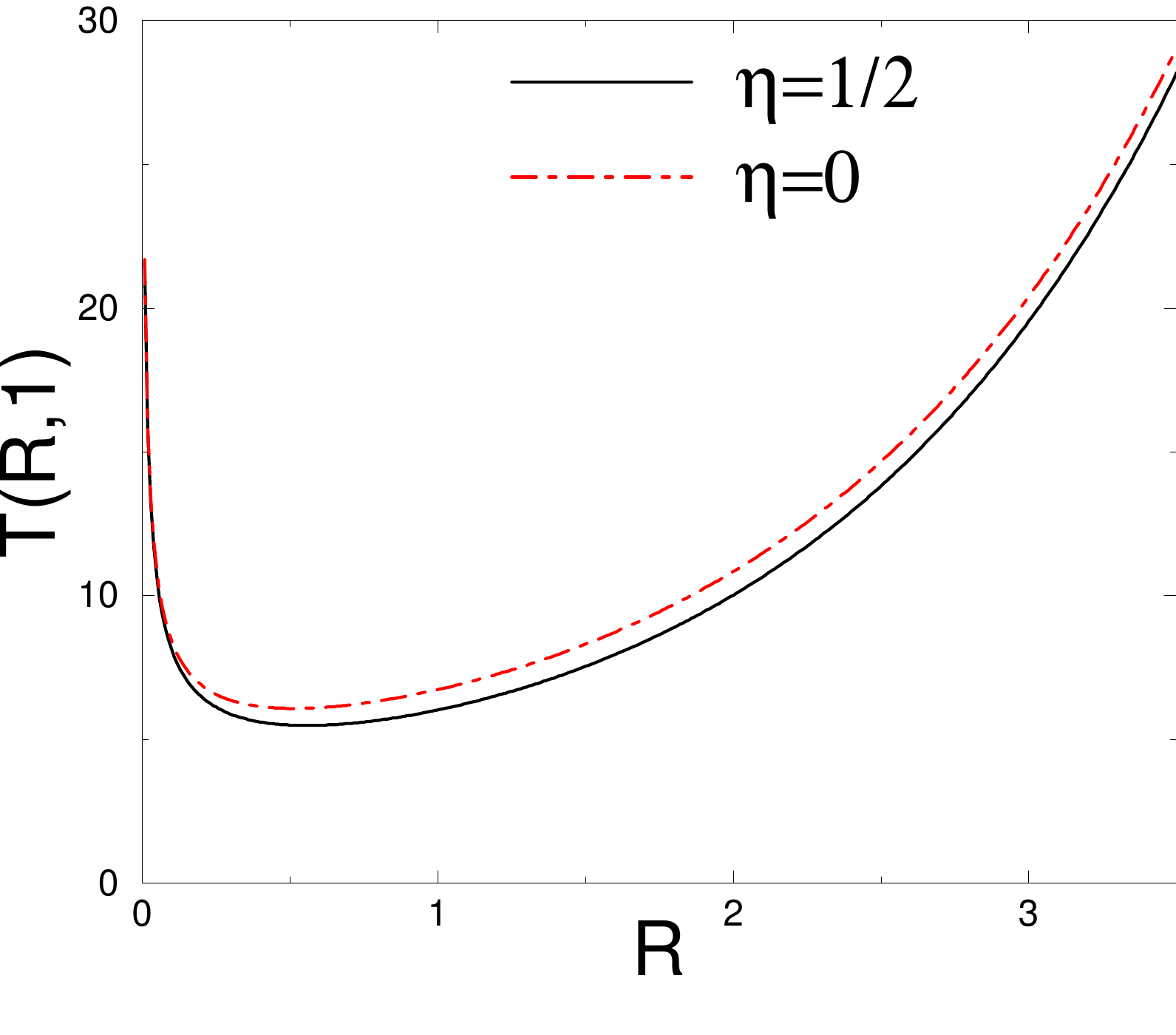}
\caption{Mean first-passage time $T(R,\xi=1)$ vs. $R$ given respectively in Eq. 
(\ref{Tred}) for $\eta=1/2$ (solid (black) line) and in Eq. (\ref{Tredeta0}) 
for $\eta=0$ (dot-dashed (red) line). For convenience, we have set the parameter
$\xi=1$ for both $\eta=1/2$ and $\eta=0$ and also $\gamma=1/2$ in both 
cases. Clearly, in both cases $T(R,1)$ has
a unique minimum and also for all $R$, $T(R,1)$ for $\eta=1/2$ is always larger
than $T(R,1)$ for $\eta=0$, indicating that the velocity randomization protocol
i.e., $\eta=1/2$ is more efficient than  position-only resetting ($\eta=0$).}
\label{fig.TR1}
\end{figure}

Let us check the diffusive limit (\ref{difflim})
in which case $\lambda \to  (r/D)^{1/2}$
and
\begin{equation}
T(X_r) \to  \frac{{\rm e}^{X_r(r/D)^{1/2}}}{r} - \frac{1}{r}
\end{equation}
recovering the result of \cite{EM11a}.

In terms of the reduced variables $R$ (\ref{Rdef}) and $\xi$ (\ref{xidef})
we obtain
\begin{eqnarray}
2\gamma T(R,\xi) &=&-\frac{1}{R} +  \frac{\left[  {\rm e}^{(R(1+R))^{1/2}\xi}
-R^2 +R^{3/2}(1+R)^{1/2} \right]}{R(1+R-(R(1+R))^{1/2})}
\label{Tredeta0}
\end{eqnarray}
which is to be compared with the $\eta =1/2$ case
(\ref{Tred}). As in the $\eta=1/2$ case, $T(R,\xi)$ as a function of $R$ for 
fixed $\xi$ has a unique minimum, signalling an optimal resetting rate (see
Fig. (\ref{fig.TR1}) for a plot). 
We see that for the same values of the reduced variables, $T(R,\xi)$ is 
always greater for  $\eta =0$  than for $\eta = 1/2$. This shows that
the velocity randomization protocol ($\eta=1/2$) is more efficient in searching
for a fixed target than  the position-only protocol ($\eta=0$), for the
same parameter values such as $X_r$, $\gamma$ and $v_0$.

It is also useful to compare the optimal search time for the run-and 
tumble dynamics with reset, with the optimal search time for the
purely diffusive search with reset. To make this comparison, we have
to set the diffusion constant $D= v_0^2/(2\gamma)$ in the diffusive
search. Setting $D=v_0^2/(2\gamma)$ in Eq. (\ref{diff_limit}) and
using $r= 2\gamma R$ and $X_r= v_0\xi/(2\gamma)$, we get
\begin{equation}
2\gamma T_{\rm diff}(R,\xi)= \frac{e^{\sqrt{R}\, \xi}-1}{R}
\label{diff_limit.2}
\end{equation}
which can be directly compared to Eq. (\ref{Tred}) for $\eta=1/2$, or with
Eq. (\ref{Tredeta0}) corresponding to $\eta=0$. Let us consider just the
$\eta=1/2$ case which is the best possibility for run and tumble dynamics.
For convenience, we set $\gamma=1/2$ and $\xi=1$ in both Eqs. 
(\ref{diff_limit.2}) and (\ref{Tred}) and optimize with respect to $R$.
For the diffusive case, we get $R^*= 2.53964\dots$ and the optimized
mean search time is then
\begin{equation}
T_{\rm diff}(R^*, \xi=1)= 1.54414\cdots
\label{diff_limit.3}
\end{equation}
In contrast, optimizing Eq. (\ref{Tred}) (the $\eta=1/2$ case), we get
$R^*= 0.55873\cdots$ (see also Fig. 1). Correspondingly, the optimized 
mean search time
is given by
\begin{equation}
T(R^*, \xi=1)=5.48571\cdots > T_{\rm diff}(R^*, \xi=1)= 1.54414\cdots
\label{compare.1}
\end{equation}
Hence, the diffusive search with reset is certainly more efficient than
the run and tumble dynamics with reset for the same set of parameters.
  
\section{Conclusion}
In this paper we have studied the resetting of a run and tumble particle in
one dimension.
First we derived the stationary state for resetting to point $X_r$ and  simultaneous velocity resetting. It turns out that the stationary state does not depend on the resetting protocol. Indeed the stationary state distribution (\ref{Pstst}) has the 
same form as a diffusive process under resetting. The width of the stationary distribution
decreases with $\gamma$ and thus increases with increasing velocity correlation time.

However the velocity resetting protocol does affect the survival probability in
the presence of an absorbing target at the origin.
We have derived explicit expressions for the mean time to absorption in the case of 
position resetting and velocity randomization (\ref{Tred}) and position-only resetting (\ref{Tredeta0}). For other parameters fixed, the position-only resetting 
gives a greater mean time to absorption.
Writing the  mean time to absorption in terms of the reduced variables $R$ (\ref{Rdef}) and $\xi$ (\ref{xidef}) we see that there is an optimal value  of $R$ which minimises the mean time to absorption.
It would be of interest to consider how these results generalise to the case of partial absorption of the particle by the boundary \cite{WEM13,Angelani15}.

Throughout we have used a renewal equation approach which facilitates the calculations.
It would be interesting to see how this approach can be extended to study the
resetting of a run and tumble particle in higher dimensions.

It would also be of interest to consider the resetting of other stochastic processes 
with correlated noise.
For example, physical Brownian motion is described as an Ornstein-Uhlenbeck process \cite{UO30}
\begin{equation}
\frac{\D x}{\D t} = v \\
\frac{\D v}{\D t} = -\gamma v + \eta(t)
\label{OU}
\end{equation}
where $\eta(t)$ is usual white noise. 
The renewal approach should again be applicable in this case.

Finally, we speculate that  progress in manipulating
bacteral swimming dynamics with light (see e.g. \cite{exp1,exp2}) may allow future experimental protocols that 
approximate to velocity resetting.

\ack
MRE acknowledges a CNRS Visiting Professorship and thanks
LPTMS for hospitality.


\section*{References}

\begin{thebibliography}{99}


\bibitem{UO30}
Uhlenbeck G E and  Ornstein L S 1930
On the Theory of the Brownian Motion
{\it Phys. Rev.} {\bf 36}, 823


\bibitem{HDL89}
Hagan P S, Doering C R,  Levermore C D 1989
The distribution of exit times for weakly colored noise
{\it J. Stat. Phys.}
{\bf 54}, 1321–1352

\bibitem{Weiss}
 Weiss G H 2002
   Some applications of persistent random walks and the telegrapher’s equation.
{\em Physica A} {\bf 311} 381

\bibitem{HV10}
S. Herrmann and P. Vallois 2010
From persistent random walk to the telegraph noise
{\it  Stoch. Dyn.} {\bf 10}, 161.




\bibitem{HLSN00}
Wu H,  Li B-L,  Springer T A,  Neill W H  2000
Modelling animal movement as a persistent random walk in two dimensions: expected magnitude of net displacement
{\it Ecological Modelling}
{\bf 132},  115-124

\bibitem{TVB12} Tejedor, V., Voituriez, R. and B\'enichou, O. 2012
 Optimizing persistent random searches. 
{\it Phys. Rev. Lett.} {\bf 108}, 088103


\bibitem{BG13}
Bhat D and Gopalakrishnan M 2013  Memory, bias, and correlations in bidirectional transport of molecular-motor-driven cargoes. 
{\it Phys. Rev. E} {\bf 88} 042702

\bibitem{Detcheverry}  Detcheverry F 2015
  Non-Poissonian  run-and-turn  motions,
{\em Europhys. Lett.}
{\bf 111} 60002


\bibitem{Rosenau}
Rosenau P 1993
Random walker and the telegrapher’s equation: A paradigm of a generalized hydrodynamics
{\it Phys. Rev. E} {\bf 48}, R655(R)



\bibitem{EG15}
Elgeti J and  Gompper G 2015, Run-and-tumble dynamics of
self-propelled particles in confinement,
{\it Europhys. Lett.}
{\bf 109} 58003 

\bibitem{Angelani17}
Angelani L 2017 
Confined run-and-tumble swimmers in one
dimension
{\it J. Phys. A: Math. Theor.}
{\bf 50}  325601.


\bibitem{SEB16}
Slowman A B, Evans M R, and Blythe R A 2016
Jamming and Attraction of Interacting Run-and-Tumble Random Walkers
{\it Phys. Rev. Lett.} {\bf 116}, 218101 

\bibitem{SEB17}
Slowman A B, Evans M R, and Blythe R A 2017
Exact solution of two interacting run-and-tumble
random walkers with finite tumble duration
{\it J. Phys. A: Math. Theor} {\bf 50}, 375601 

\bibitem{Angelani15}  Angelani L 2015. Run-and-tumble particles, telegrapher's equation and absorption problems
with partially reflecting boundaries. 
{\it J. Phys. A: Math. Theor.}
{\bf 48} 495003.


\bibitem{MJKKSMRD18}
Malakar K
et al 2018
Steady state, relaxation and first-passage
properties of a run-and-tumble particle in one-
dimension
{\it J. Stat. Mech.}  043215

\bibitem{EM11a} Evans M R and Majumdar S N 2011
 Diffusion with stochastic resetting,
 {\it  Phys. Rev. Lett.}  {\bf 106}, 160601 

\bibitem{MZ99} Manrubia S C and Zanette D H 1999 Stochastic multiplicative processes  with reset events, {\it Phys. Rev. E} {\bf 59}, 4945


\bibitem{MMV17}
Montero M,  Mas\'o-Puigdellosas A, Villarroel J 2017 
Continuous-time random walks with reset events: Historical background and new perspectives
{\it European Physical Journal B}
{\bf 90} 176 




\bibitem{EM11b} Evans M R and Majumdar S N 2011 
 Diffusion with optimal resetting,
 {\it  J. Phys. A: Math. Theor.}  {\bf 44}, 435001 

\bibitem{EM14}
Evans M R and  Majumdar S N 2014
Diffusion with resetting in arbitrary spatial dimension
{\it  J. Phys. A: Math. Theor.} {\bf 47},  285001  

\bibitem{MSS15}
 Majumdar S N, Sabhapandit S and Schehr G 2015,
Dynamical transition in the temporal relaxation of stochastic processes under resetting
{\it Phys. Rev. E} {\bf 91} 052131 

\bibitem{BenichouRV}   B\'enichou O,  Loverdo C, Moreau M, and Voituriez R 2011,
    Intermittent search strategies, {\it Rev. Mod. Phys.}{ \bf 83}, 81. 


\bibitem{MZ02}  Montanari A and  Zecchina R 2002,
Optimizing searches via rare events,
{\it Phys. Rev. Lett.} {\bf 88}, 178701 


\bibitem{EM16}
Eule S and Metzger J J  2016
Non-equilibrium steady states of stochastic processes with intermittent 
resetting
{\it  New J. Phys.}  {\bf 18},  033006  

\bibitem{PKE16}
Pal A, Kundu A and Evans M R 2016
Diffusion under time-dependent resetting
{\it  J. Phys. A: Math. Theor.} {\bf 49},  225001  

\bibitem{NG16}
 Nagar A and Gupta S 2016
Diffusion with stochastic resetting at power-law times
{\it  Phys. Rev. E}  {\bf 93},  060102 (R)  

\bibitem{BDR16}
Bhat U, De Bacco C, Redner S 2016 
Stochastic Search with Poisson and Deterministic Resetting
{\it J. Stat. Mech.}  P083401  

\bibitem{HK16}
Husain K and Krishna S 2016
Efficiency of a Stochastic Search with Punctual and Costly Restarts
{\em Preprint} arXiv: 1609.03754

\bibitem{PR17}  Pal A and Reuveni S 2017  First Passage under Restart
{\it Phys. Rev. Lett.}
{\bf 118}  030603


\bibitem{BS2014}
Boyer D and Solis-Salas C 2014
Random walks with preferential relocations to places visited in the past
and their application to biology {\it Phys. Rev. Lett.} {\bf 112}, 240601

\bibitem{MSS15b} Majumdar S~N, Sabhapandit S and Schehr G 2015
Random walk with random resetting to the maximum position
{\it Phys. Rev. E} {\bf 92},  052126

\bibitem{BEM17} Boyer D, Evans M R and Majumdar S N 2017
Long time scaling behaviour for diffusion with resetting and memory
{\it J. Stat. Mech.} {\bf P023208}

\bibitem{FBGM17} Falcon-Cortes A, Boyer D, Giuggioli L, and Majumdar S N
2017 Localization transition induced by learning in random searches
{\it Phys. Rev. Lett.} {\bf 119}, 140603    


\bibitem{EMM13}
Evans M R, Majumdar S N and Mallick K 2013
Optimal diffusive search: nonequilibrium resetting versus equilibrium dynamics
{\it J. Phys. A: Math. Theor}
{\bf 46}  185001


\bibitem{R16} Reuveni S 2016 Optimal stochastic restart renders fluctuations
in first-passage times universal {\it Phys. Rev. Lett.} {\bf 116} 170601


\bibitem{BMS13} Bray A J, Majumdar S N and Schehr G 2013 Persistence and first-passage 
properties in nonequilibrium systems {\it Adv. in Phys.} {\bf 62} 225 

\bibitem{KMSS14} Kusmierz L, Majumdar S N, Sabhapandit S and Schehr G 2014
First order transition for the optimal search time of L\'evy flights with 
resetting {\it Phys. Rev. Lett.} {\bf 113} 220602

\bibitem{KG15} Kusmierz L and Gudowska-Nowak E 2015 Optimal first-arrival 
times in L\'evy flights with restting {\it Phys. Rev. E} {\bf 92} 052127

\bibitem{CM15} Campos D and M\'endez V 2015
Phase transitions in optimal
search times: How random walkers should combine reset-
ting and flight scales {\it Phys. Rev. E} {\bf 92} 062115 

\bibitem{CS15} Christou C and Schadschneider A 2015 Diffusion with resetting
in bounded domain {\it
J. Phys. A: Math. Theor.} {\bf 48} 285003 

\bibitem{GMS13} Gupta S, Majumdar S N and  Schehr G 2014
Fluctuating interfaces subject to stochastic resetting,
{\it Phys. Rev. Lett.} {\bf 112} 220601 

\bibitem{DHP14}
Durang X, Henkel M, Park H 2014
Statistical mechanics of the coagulation-diffusion process with a stochastic reset
{\it  J. Phys. A: Math. Theor.} {\bf 47}  045002 

\bibitem{RRU15} Rothart T,
Reuveni S and Urbakh M 2015 Michaelis-Menten reaction scheme as a unified
approach towards the optimal restart problem
{\it Phys. Rev. E}  {\bf 92}, 060101

\bibitem{FGS16}
Fuchs J, Goldt S and Seifert U 2016
Stochastic thermodynamics of resetting
{\it EPL }  {\bf 113}  60009

\bibitem{PRahav17} Pal A and Rahav S 2017 Integral fluctuation theorems
for stochastic resetting systems {\it Phys. Rev. E} {\bf 96} 062135
 

\bibitem{MST15} Meylahn J M, Sabhapandit S, and Touchette H 2015
Large deviations of Markov processes with resetting {\it Phys. Rev. E}
{\bf 92} 062148 

\bibitem{HT17}
Harris R J and Touchette H 2017
Phase transitions in large deviations of reset processes
{\it J. Phys. A: Math. Theor.} {\bf 50} 10LT01 

\bibitem{HMMT18} Hollander H D, Majumdar SN, Meylahn J M, and
Touchette H 2018 Properties of additive functionals of Brownian motion 
with resetting {\it preprint} arXiv:1801.09909 

\bibitem{FE17}
Falcao R and Evans M R 2017
Interacting Brownian motion with resetting
{\it J. Stat. Mech.} 023204

\bibitem{PER18} Pal A, Eliazar I,  Reuveni S 2018 
First passage under restart with branching
{\em preprint} arXiv:1807.09363

\bibitem{MO18}  Majumdar S N and Oshanin G 2018
Spectral content of fractional Brownian motion with stochastic reset
{\it J. Phys. A: Math. Theor.} {\bf 51} 435001 

\bibitem{MSM18} Mukherjee B, Sengupta K and Majumdar SN 2018
Quantum dynamics with stochastic reset 
{\em preprint} arXiv:1806.00019

\bibitem{RTLG18} Rose D C, Touchette H, Lesanovsky I and Garrhan J P 2018
Spectral properties of simple classical and quantum reset processes
{\it Phys. Rev. E}  {\bf 98}, 022129 




\bibitem{CS18}
Chechkin A and  Sokolov I M 2018
Random Search with Resetting: A Unified Renewal Approach
{\it Phys. Rev. Lett.} {\bf 121} 050601 


\bibitem{Othmer} Othmer H G, Dunbar S R, and Alt W 1988
Models of dispersal in biological systems {\it J. Math. Biol.}
{\bf 26} 263

\bibitem{Martens} Martens K, Angelani l, Di Leonardo R, and Bocquet L 2012
Probability distributions for the run-and-tumble bacterial dynamics: An 
analogy to the Lorentz model {\it Eur. Phys. J. E} {\bf 35}, 84


\bibitem{WEM13}
Whitehouse J, Evans M R, and  Majumdar S N 2013
Effect of partial absorption on diffusion with resetting
{\it Phys. Rev. E}  {\bf 87}, 022118 

\bibitem{exp1}
Walter J M, Greenfield D,  Bustamante C and Liphardt J 2007
Light-powering Escherichia coli with proteorhodopsin
{\em PNAS} {\bf 104}, 2408

\bibitem{exp2}
Arlt J, Martinez V A, Dawson A, Pilizota T and Poon WCK 2018
Painting with light-powered bacteria
{\it Nat. Commun.} {\bf 9}, 768.










\end{thebibliography}
\end{document}